\documentclass[12pt,twoside,a4paper]{article}
\usepackage{amsmath}
\usepackage{epsfig}

\font\bit=cmbxti10 
\def \osix {\hbox{$\alpha_s \langle \bar{q}q \rangle^2$}}

\begin{document}

\title{ 
\vskip-3cm 
{\normalsize
\hfill MZ-TH/03-21\\
\hfill December 2003 \\
}
\vskip4cm 
{\bf QCD condensates from $\tau$-decay data:}\\  
{\bf A functional approach}\thanks{Supported by PROCOPE, EGIDE} 
\\[3ex] 
}

\date{}
\author{
S.~Ciulli, 
C.~Sebu\footnote{Email: sebu@lpm.univ-montp2.fr} \\[1ex]
\normalsize{Laboratoire de Physique-Math\'ematique et Th\'eorique,}\\
\normalsize{Universit\'e de Montpellier II, 34095 Montpellier, France} 
\\[3ex] 
K.~Schilcher, 
H.~Spiesberger\footnote{Email: hspiesb@thep.physik.uni-mainz.de} \\[1ex]
\normalsize{Institut f\"ur Physik, Johannes-Gutenberg-Universit\"at,}\\ 
\normalsize{Staudinger Weg 7, D-55099 Mainz, Germany} \\[10ex]
}

\maketitle

\begin{abstract}
\medskip 
\noindent 
We study a functional method to extract the $V-A$ condensate of
dimension 6 from a comparison of $\tau$-decay data with the asymptotic
space-like QCD prediction. Our result is in agreement within errors with
that from conventional analyses based on finite energy sum rules.
\end{abstract}

\thispagestyle{empty}

\clearpage  

\section{Introduction}

Although QCD has been with us for three decades, the knowledge of the
values of the various fundamental or effective parameters of the theory
(with the possible exception of the coupling constant) such as quark
masses and condensates is still astonishingly limited. The precise data
on $\tau $-decay obtained by the ALEPH \cite{aleph} and OPAL \cite{opal}
collaborations at CERN have offered an opportunity for new studies,
which range from an extraction of the strange quark mass
\cite{quarkmasses} to the determination of various condensate
parameters. Of particular interest was the extraction of the dimension-6
condensate \cite{DGHS,BGP,IZ,CAD+KS,Golo} which, in the chiral limit,
determines, e.g., the $K\rightarrow \pi \pi$ matrix elements of the
relevant electroweak penguin operators.

It should be kept in mind, however, that the extraction of the
condensate parameters of QCD constitute a so-called ill-posed inverse
problem which is basically unstable with respect to errors in the input
data. An example is the popular method used to obtain condensates from
experimental spectral functions by QCD sum rules (finite-energy sum
rules, FESR). This procedure corresponds to an analytic continuation
from a finite contour (the part of the positive real axis on which the
data are given) which is known to be notoriously unstable with respect
to data errors. The extraction of QCD condensates requires therefore a
carefully chosen stabilization mechanism. In the case of QCD sum rules,
this is achieved by an implicit and rather {\it ad hoc} assumption that
the series of the operator product expansion essentially breaks off
after a finite number of terms.  Amazingly, a careful analysis of finite
energy sum rules \cite{CAD+KS} of the chiral spectral function shows
remarkable stability with respect to variations of the duality interval.
To investigate further the reliability of the extraction of QCD
parameters via sum rules, it would be prudent to develop alternative
methods.

In this letter we study a functional method \cite{auberson1,auberson2}
which allows us to extract without prejudice the condensates from a
comparison of the time-like data with the asymptotic space-like QCD
results. We will see that the price to be paid for the increased
credibility are possibly larger errors in the values of the extracted
parameters.


\section{QCD condensates}

We consider the polarization operator of hadronic vector and axial-vector
charged currents, $J_{\mu }=V_{\mu }=\bar{u}\gamma _{\mu }d$ and $J_{\mu
}=A_{\mu }=\bar{u}\gamma _{\mu }\gamma _{5}d$, 
\begin{eqnarray}
\Pi _{\mu \nu }^{J} 
& = & i\int dxe^{iqx}\langle TJ_{\mu }(x)J_{\nu}(0)^{\dagger}\rangle 
\\
& = & \left( -g_{\mu \nu } q^{2} + q_{\mu }q_{\nu } \right) 
\Pi_{J}^{(1)}(q^{2}) + q_{\mu }q_{\nu }\Pi _{J}^{(0)}(q^{2}) \,.
\nonumber
\end{eqnarray}
The conservation of the vector current implies $\Pi _{V}^{(0)}=0$.

The spectral functions are related to the absorptive part of the
correlators
\begin{equation}
v_{j}(s) = 4\pi {\rm Im}\Pi _{V}^{(j)}(s),~~~~a_{j}(s)
         = 4\pi {\rm Im}\Pi_{A}^{(j)}(s) 
\end{equation}
and can be measured in hadronic $\tau $-decays. We consider specifically
the $V-A$ component which is related to the branching ratios of $\tau$
decays through
\begin{eqnarray}
& & R_{\tau, V-A} = \frac{B(\tau \rightarrow \nu _{\tau }+{\rm
    hadrons},~V-A)}{B(\tau \rightarrow \nu _{\tau }+e+\bar{\nu}_{\tau })}
\label{rtauvma} 
\\
& = & 6\left\vert V_{ud}\right\vert ^{2}S_{{\rm EW}}
\int_{0}^{m_{\tau}^{2}} \frac{ds}{m_{\tau }^{2}}
\left(1-\frac{s}{m_{\tau}^{2}}\right)^{2} 
\left[ \left(1 + 2\frac{s}{m_{\tau }^{2}}\right) 
\left( v_{1}-a_{1}-a_{0}\right) + \frac{2}{m_{\tau }^{2}}sa_{0}\right]
\,.  
\nonumber
\end{eqnarray}
Here, $V_{ud}$ is the weak mixing CKM-matrix element, $\left| V_{ud}
\right|^2 = 0.9752 \pm 0.0007$, the $\tau$ mass is denoted $m_{\tau} =
1.777$ GeV and $S_{\rm EW} = 1.0194 \pm 0.0040$ accounts for electroweak
radiative corrections \cite{sew}.  The spin-0 axial vector contribution
$a_{0}(s)$ is dominated by the one-pion state, $a_{0}(s)=2\pi ^{2}f_{\pi
  }^{2}\delta (s-m_{\pi }^{2})$, with the $\pi$-decay constant $f_{\pi}
= 0.1307$ GeV.  Its contribution in the last term of (\ref{rtauvma}) is
tiny
\begin{equation}
\left. 
\Delta R_{\tau ,V-A} \right\vert_{a_{0}} \simeq 24\pi^{2}
\frac{f_{\pi}^{2}m_{\pi}^{2}}{m_{\tau}^{4}} \simeq 0.0074 
\end{equation}
and will be neglected. The contribution of the pion pole to the first
term in Eq.\ (\ref{rtauvma}) is well identified in the data and
concentrated at low $s \simeq m_{\pi}^2$; thus it can be removed from
the data and taken into account explicitly without introducing sizable
additional uncertainties. The experimental data
\cite{aleph,opal}\footnote{We use the ALEPH data \cite{aleph} because of
  their smaller experimental errors.} are given by binned and normalized
event numbers related to the differential distribution
$dR_{\tau,V-A}/ds$ and can therefore be viewed as a measurement of the
function
\begin{equation}
\omega_{V-A}(s) = v_{1}(s)-a_{1}(s)-a_{0}(s). 
\end{equation}

The $(V-A)$ correlator is special since it vanishes identically in the
chiral limit ($m_{q}=0)$ to all orders in QCD perturbation theory.
Renormalon ambiguities are thus avoided. Non-perturbative terms can be
calculated for large $\left\vert s\right\vert $ by making use of the
operator product expansion (OPE) of QCD
\begin{equation}
\Pi_{V-A}^{(0+1)}(s) = \sum_{D\geq 4} 
\frac{O_{D}^{V-A}}{(-s)^{D/2}}
\left(1 + c_{D}\frac{\alpha _{s}}{\pi }\right)
\label{ope}
\end{equation}
where $O_{D}^{V-A}$ are vacuum matrix elements of local operators of
dimension $D$ (so-called condensates). Their contribution is known up to
dimension 8 and read, at leading order,
\begin{eqnarray}
O_{4}^{V-A} & = & (m_{u}+m_{d}) \langle \bar{q}q\rangle 
= -f_{\pi}^{2}m_{\pi}^{2} \, , 
\\ 
O_{6}^{V-A} & = & -\frac{32\pi}{9}\alpha_{s}\langle \bar{q}q\rangle^2 
\, ,
\\
O_{8}^{V-A} & = & 4\pi \alpha_{s} i\langle \bar{q} G_{\alpha \beta}
G^{\alpha \beta} q \rangle \, .
\label{opeo8}
\end{eqnarray}
The last two results hold in the vacuum dominance approximation.  The
numerical value of $O_{4}^{V-A}$ is very small and this condensate can
be neglected in our analysis. The next-to-leading-order corrections to
$O_{6}^{V-A}$ have been calculated in \cite{nlo1,nlo2}.  They depend on
the regularization scheme implying that the value of the condensate
itself is a scheme-dependent quantity. Explicitly,
\begin{equation}
O_6^{V-A} = -\frac{32\pi}{9}\alpha_s \langle \bar{q}q\rangle^{2}
\left(1 + \frac{\alpha_s(\mu^2)}{4\pi} 
          \left[c_6 + \ln \left(\frac{\mu^2}{-s}\right) \right] 
\right) 
\label{pivma}
\end{equation}
where 
\begin{equation}
c_{6}=\left\{ 
\begin{array}{ll}
\displaystyle \frac{247}{48} & 
              {\rm BM-scheme~\cite{nlo1},} \\[2ex] 
\displaystyle \frac{89}{48}  & 
              {\rm anticommuting}~\gamma _{5}~{\rm \cite{nlo2}.}
\end{array}
\right. 
\label{c6}
\end{equation}
The renormalization scale $\mu ^{2}$ is conveniently chosen to be $-s$.

The result for $O_{8}^{V-A}$, Eq.\ (\ref{opeo8}), is taken from
\cite{IZ}. It involves a quark-gluon condensate for which various
estimates exist. The typical scales determining the condensates are
around 300 MeV, e.g.\ $\langle \bar{q}q\rangle \simeq (250 \, {\rm
  MeV})^{3}$, ($\alpha _{s}/\pi )\langle G^{2}\rangle \simeq (300 \,
{\rm MeV})^{3}$. Assuming a similar scale for the condensate entering
$O_8^{V-A}$, we expect $O_8^{V-A}$ to be of order $10^{-3}$ GeV$^8$.
This is small enough so that the OPE makes sense.  If $O_8^{V-A}$ would
be larger, radiative corrections to higher-dimension condensates would
mix significantly with the lower-dimension condensates through their
imaginary parts.  There exist a number of QCD sum rule extractions of
the value of the $D=8$ condensate. They range from
($-7.5^{+5.2}_{-4.0})\cdot 10^{-3}$ GeV$^8$ \cite{BGP} to $(4.4\pm
1.2)\cdot 10^{-3}$ GeV$^8$ \cite{DGHS}. A recent conservative estimate
\cite{CAD+KS} is $O_8 = (-1.0 \pm 6.0)\cdot 10^{-3}$ GeV$^8$.  This
value corresponds to a scale of about 400 MeV which is comparable to
$\Lambda_{\rm QCD}$.  The variation of these results represents the
ambiguities inherent in the QCD sum rule approach.  In the next section we
shall present our alternative rigorous functional method which allows
us to extract the condensates from a comparison of the data with the
asymptotic space-like QCD results in an unambiguous way.


\section{An $L^2$ norm approach}

We consider a set of functions $F(s)$ (where $F(s)$ relates to
$\Pi_{V-A}^{(0+1)}(s)$) expressed in terms of some squared energy
variable $s$ which are admissible as a representation of the true
amplitude if
\begin{itemize}
\item[\bit i)] $F(s)$ is a real analytic function in the complex
  $s$-plane cut along the time-like interval $\Gamma _{R}=[s_{0},\infty
  )$. The value of the threshold $s_{0}$ depends on the specific
  physical application ($s_{0}=(2m_{\pi })^{2}$ for $\Pi_{V}$,
  $s_{0}=m_{\pi }^{2}$ for $\Pi_{A}$).
\item[\bit ii)] The asymptotic behavior of $F(s)$ is restricted by
  fixing the number of subtractions in the dispersion relation between
  $F(s)$ and its imaginary part along the cut $f(s)={\rm
    Im}F(s+i0)|_{s\in \Gamma _{R}}$ (for $\Pi_{V-A}^{(0+1)}(s)$ no
  subtractions are needed):
\begin{equation}
F(s)=\frac{1}{\pi }\int_{s_{0}}^{\infty }\frac{f(x)}{x-s}dx\,.
\label{disprel}
\end{equation}
\end{itemize}

We have two sources of information which will be used to determine
$F(s)$ and $f(s)$. First, there are experimental data in a {\sl
  time-like interval} $\Gamma_{\rm exp} = [s_0, s_{\rm max}]$ with $s_0
> 0$ for the imaginary part of the amplitude. Although these data are
given on a sequence of adjacent bins, we describe them by a function
$f_{{\rm exp}}(s)$. We assert that $f_{{\rm exp}} $ is a real, not
necessarily continuous function. The experimental precision of the data
is described by a covariance matrix $V(s, s^{\prime})$.

On the other hand, we have a theoretical model, in fact QCD. From
perturbative QCD we can obtain a prediction for the amplitude in a {\sl
  \ space-like interval}\footnote{We do not exclude the case $s_2
  \rightarrow -\infty$.} $\Gamma_L = [s_2, s_1] $. This model amplitude
$F_{{\rm QCD}}(s)$ is a continuous function of real type, but does not
necessarily conform to the analyticity property {\bit i)}. Since
perturbative QCD is expected to be reliable for large energies, we
expect that there is also useful information about the imaginary part of
the amplitude provided that $|s|$ is large, i.e.\ we can also use
$f_{{\rm QCD}}(s) = {\rm Im} F_{{\rm QCD}}(s+i0)|_{s \in (s_{\rm
    max},\infty)}$. In order to compare the true amplitude with theory,
we can therefore split the integral in the dispersion relation
(\ref{disprel}),
\begin{equation}
F(s)
-\frac{1}{\pi }\int_{s_{\rm max}}^{\infty }\frac{f(x)}{x-s}dx 
=\frac{1}{\pi }\int_{s_{0}}^{s_{\rm max} }\frac{f(x)}{x-s}dx \, ,
\label{disprel1}
\end{equation}
and test the hypothesis whether the left-hand side can be described by
QCD. 

We also need an {\em a-priori} estimate of the accuracy of the QCD
predictions. This will be described by a continuous, strictly positive
function $\sigma_{L}(s)$ for $s \in \Gamma_{L}$ which should describe
errors due to the truncation of the perturbative series and the operator
product expansion and is expected to decrease as $|s|\rightarrow \infty
$ and diverge for $s\rightarrow 0$. In the case of $\Pi_{V-A}$ which
does not have perturbative contributions, we will take the contribution
of the dimension $D=8$ operator as an error and use $\sigma_{L}(s) =
O_{8}/s^{4}$ with $O_{8}$ in the order of $10^{-3} \, {\rm GeV}^{8}$. If
the perturbative part dominates, as is the case for the individual
vector or axial vector correlators, the last known term of the
perturbation series could be used as a sensible estimate of the error
corridor.

The goal is to check whether there exists any function $F(s)$ with the
above analyticity properties, the true amplitude, which is in accord
with both the data in $\Gamma _{{\rm exp}}$ and the QCD model in
$\Gamma_{L}$. In order to quantify the agreement we will define
functionals $\chi _{L}^{2}[f]$ and $\chi _{R}^{2}[f]$ using an $L^{2}$
norm. For the time-like interval we simply compare the true amplitude
$f(s)$ with the data and use the covariance matrix of the experimental
data as a weight function:
\begin{equation}
\chi_{R}^{2}[f] = \int_{s_{0}}^{s_{\rm max}} dx 
\int_{s_{0}}^{s_{\rm max}} dx^{\prime}\,
V^{-1}(x,x^{\prime }) \left(f(x)-f_{{\rm exp}}(x)\right) 
  \left( f(x^{\prime})-f_{{\rm exp}}(x^{\prime })\right) \,.
\label{chir2}
\end{equation}
Experimental data correspond to cross sections measured in bins of $s$,
so that we can calculate this integral in terms of a sum over data
points. The ALEPH data which we use are given for 65 equal-sized bins of
width $\Delta s = 0.05$ GeV$^2$ between $0$ and $3.25$ GeV$^2$.
$\chi_R^2$ given in (\ref{chir2}) is in fact the conventional definition
of a $\chi^{2}$ and has a probabilistic interpretation: for uncorrelated
data obeying a Gaussian distribution we would expect to obtain
$\chi_{R}^{2}=N$, where $N$ is the number of data points. Since
experimental data at different energies are correlated, we instead
expect
\begin{equation}
\chi_{{\rm exp}}^{2} = 
\sum_{i,j} \sqrt{V(s_{i},s_{i}) V(s_{j},s_{j})}
V^{-1}(s_{i},s_{j}) \, .
\end{equation}

In order to define a measure for the agreement of the true function
$f(s)$ with theory, we use the left-hand side of (\ref{disprel1}) which
is well-defined and expected to be a reliable prediction of QCD in the
space-like interval for not too small $|s|$. This expression can be
compared with the corresponding integral over the true function. Thus we
define
\begin{equation}
\chi^2_L[f] = N \int_{\Gamma_L} w_L(x) \left( F_{{\rm QCD}}(x) 
- \frac{1}{\pi}\int_{s_{\rm max}}^{\infty} 
\frac{f_{{\rm QCD}}(x^{\prime})}{x^{\prime}-x}dx^{\prime} 
- \frac{1}{\pi} \int_{s_0}^{s_{\rm max}} 
\frac{f(x^{\prime})}{x^{\prime}-x}dx^{\prime}\right)^2 dx
\label{chil2}
\end{equation}
where $w_L$ is the weight function for the space-like interval and
identified with $1/\sigma_L^2(s)$. The integral is normalized to unity
for the case where the difference within parentheses saturates the error
$\sigma_L$. 

In order to find the true function $f(s)$, we can combine the
information contained in $\chi_R^2$, (\ref{chir2}), and $\chi_L^2$,
(\ref{chil2}) in the following way \cite{auberson1,auberson2}. We fix
\begin{equation}
\chi^2_R[f] = \chi^2_0 \leq \chi^2_{{\rm exp}} \, ,
\end{equation}
and minimize $\chi^2_L$: 
\begin{equation}
\chi^2_L[f] \rightarrow least \quad (\equiv \chi^2_{L,{\rm min}}) \, .
\end{equation}
These conditions are equivalent to finding the unrestricted minimum of the
functional 
\[
{\cal F} [f] = \chi^2_L[f] + \mu \chi^2_R[f] 
\]
where $\mu$ is the Lagrange multiplier, which will be found later. The
solution of the condition $\chi^2_R[f] = \chi^2_0$ will be denoted by
$f(x;\mu)$:
\[
\int_{s_0}^{s_{\rm max}} dx \int_{s_0}^{s_{\rm max}} dx^{\prime}
V^{-1}(x,x^{\prime}) \left[f(x;\mu) - f_{{\rm exp}}(x)\right] 
\left[f(x^{\prime};\mu) - f_{{\rm exp}}(x^{\prime})\right] = \chi^2_0 \, . 
\]
To this end we require the Fr{\'e}chet derivative of ${\cal F}$ to be zero 
\[
\partial {\cal F} [f, Y] \equiv \lim_{\alpha \rightarrow 0} \frac{\partial 
{\cal F} [f+\alpha Y]}{\partial \alpha} = 0 \, , 
\]
for any function $Y$. This leads to the following integral equation for
the imaginary part $f(x;\mu)$: 
\begin{eqnarray}
f(x;\mu) 
= f_{{\rm exp}}(x) 
& + & \frac{\lambda}{\pi} 
      \int_{s_0}^{s_{\rm max}} dx^{\prime}V(x,x^{\prime}) 
      \int_{\Gamma_L} dx^{\prime\prime}
                      w_L(x^{\prime\prime}) F_{{\rm QCD}}(x^{\prime\prime}) 
                      \frac{1}{x^{\prime}-x^{\prime\prime}}  
\nonumber 
\\
& - & \frac{\lambda}{\pi^2} 
      \int_{s_0}^{s_{\rm max}} dx^{\prime}V(x,x^{\prime})
      \int_{s_{\rm max}}^{\infty} dx^{\prime\prime}
      \int_{\Gamma_L} dy w_L(y) 
\frac{f_{{\rm QCD}}(x^{\prime\prime})}{(x^{\prime}-y)(x^{\prime\prime}-y)}  
\nonumber \\
& + & 
\lambda \int_{s_0}^{s_{\rm max}} dx^{\prime}
            {\cal K}(x,x^{\prime})f(x^{\prime};\mu)\, ,  
\label{fsolv}
\end{eqnarray}
where $\lambda = 1/\mu$ and 
\[
{\cal K}(x,x^{\prime}) = - \frac{1}{{\pi}^2} \int_{s_0}^{s_{\rm max}}
dx^{\prime\prime}V(x,x^{\prime\prime}) \int_{\Gamma_L} dy 
\frac{w_L(y)}{(x^{\prime}-y)(x^{\prime\prime}-y)} \, . 
\]
Eq.\ (\ref{fsolv}) is a Fredholm equation of the second type which is
stable against variations of its input. At this stage we should notice
that if one had claimed that in the space-like region the function
$F(s)$ was given by some analytic expression (e.g., by some few QCD
terms), this would be equivalent to saying that $\chi_L^2$ vanished
identically. But then, from the definition of the functional ${\cal F}$
and the vanishing of its Fr{\'e}chet derivative, it follows that $\mu =
1/\lambda$ is zero which will turn the integral equation (\ref{fsolv})
into a Fredholm equation of the first kind which is known to be
unstable.

The integral equation will be solved numerically by expanding $f(s)$ in
terms of Legendre polynomials.  The algorithm to determine an acceptable
value for the condensate is then the following:
\begin{itemize}
\item[\bit i)] For a fixed value of $\chi_0^2 = \chi^2_{\rm exp}$ we
  determine the solution (\ref{fsolv}) and calculate the corresponding
  value of $\chi_L^2[f]$ as a function of the condensate $\osix$. The
  Lagrange multiplier $\mu$ is determined by iteration such that the
  condition $\chi_R^2[f]=\chi_0^2$ is fulfilled. 
  
\item[\bit ii)] We minimize this $\chi_L^2[f]$ with respect to $\osix$
  and call the minimal value $\chi_{L, {\rm min}}^2$ and the
  corresponding $\alpha_s \langle \bar{q} q\rangle^2$ is the value for
  the condensate we are looking for.
  
\item[\bit iii)] We determine the error on $\osix$ by solving $\chi
  _{L}^{2}(\osix)=\chi _{L,{\rm min}}^{2}+1$.
\end{itemize}


\section{Numerical results and discussion}

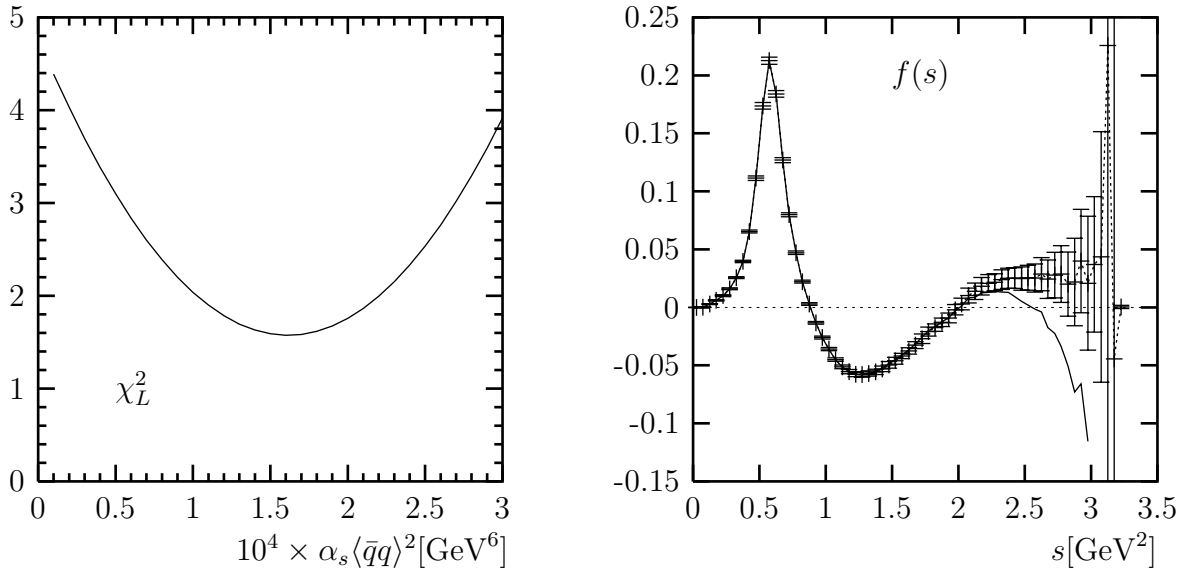
\begin{figure}[th!]
\unitlength 1mm 
\begin{picture}(156,72)
\put(-2,-5){\begin{minipage}[b][68mm][b]{68mm}
\include{chit2s0}
\end{minipage}}
\put(78,-5){\begin{minipage}[b][68mm][b]{68mm}
\include{frt2s0}
\end{minipage}}
\end{picture}
\caption{A typical result for $\protect\chi_L^2$ as a function of
  $\osix$ (left) and the regularized function compared with data
  \protect\cite{aleph} (right). We have chosen $O_8 = 10^{-3}$ GeV$^8$
  and $c_6 = 89/48$.}
\label{fig1}
\end{figure}

A typical situation resulting from this algorithm is shown in Fig.\ 
\ref{fig1}. The left part of this figure shows $\chi_L^2$ which has the
expected quadratic dependence of $\osix$. The values of $\osix$
corresponding to the minimum of $\chi_L^2$ are listed in the Table below
for various choices of $O_8$ as discussed above.  The regularized
function shown in the right part of Fig.\ \ref{fig1} follows nicely the
data points, except at large $s$. Here the experimental errors are large
and hence, as it should happen, the regularizing effect by means of the
functional (\ref{chir2}) is not as effective.

In the numerical evaluation we have used the NLO expression for
$\alpha_s(s)$ with $\Lambda_{\overline{{\rm MS}}}^{N_f=3} = 0.326$ GeV.
The result for $\osix$ is not sensitive to changing
$\Lambda_{\overline{{\rm MS}}}^{N_f=3}$ within the present experimental
error $\pm 0.030$ GeV. For the evaluation of $\chi_L^2$ we have
restricted the range of integration within limits $s_2 \le s \le s_1 <
0$. We checked that our result is insensitive to changes of $s_2$ as
soon as its absolute value is chosen larger than $O(100)$ GeV$^2$. Since
the error channel defined by $O_8/s^4$ diverges for $s \rightarrow 0$,
one could, in principle, choose the upper limit $s_1 = 0$. Numerical
instabilities require a non-zero value. We observe a well-defined
plateau for the result for $\osix$ as a function of $s_1$ between $-1.0$
and $-0.5$ GeV$^2$ and quote the values for $s_1 = -0.7$ GeV$^2$.

\begin{table}[hb!]
\begin{center}
\begin{tabular}{rcc}
\hline\hline
\rule[0mm]{0mm}{5mm} 
    & $\osix$ for $c_6 = \frac{89}{48}$  
        & $\osix$ for $c_6 = \frac{247}{48}$ 
\\[1mm] \hline
\rule[0mm]{0mm}{5mm} $O_8 = 1.0 \times 10^{-3}$ GeV$^8$ 
    & $1.6 \pm 1.0$  & $1.1 \pm 0.6$ 
\\[1mm] 
\rule[0mm]{0mm}{5mm} $1.25 \times 10^{-3}$ GeV$^8$ 
    & $1.6 \pm 1.1$  & $1.1 \pm 0.8$ 
\\[1mm] 
\rule[0mm]{0mm}{5mm} $1.5 \times 10^{-3}$ GeV$^8$ 
    & $1.6 \pm 1.2$  & $1.1 \pm 0.9$ 
\\[1mm] \hline\hline
\end{tabular}
\end{center}
\caption{Results of the determination of $\osix$ (in units of $10^{-4}$
  GeV$^6$) for the two choices of $c_6$ in (\protect\ref{c6}) and with
  different values for $O_8$ to fix the error channel in the space-like
  interval.} 
\end{table}

The values for $\osix$ given in the table translate into values for the
condensate ${\cal O}_6^{V-A}$ according to Eq.\ (\ref{pivma}). In order
to compare with other results from the literature we use $\alpha_s(s) =
0.6$ at the scale $s = 1$ GeV$^2$. For $O_8 = 1.0 \times 10^{-3}$
GeV$^8$ we obtain
\begin{equation*}
{\cal O}_6^{V-A} = 
\left\{
\begin{array}{ll}
\displaystyle
(-0.0020 \pm 0.0014)~{\rm GeV}^{6} & {\rm for}~~ c_6 = \frac{89}{48}\, ,
\\[1ex]
\displaystyle
(-0.0015 \pm 0.0009)~{\rm GeV}^{6} & {\rm for}~~ c_6 = \frac{247}{48}\, .
\end{array}
\right.
\end{equation*}
These results can be compared with the lowest-order vacuum saturation
expression
\begin{equation*}
\left. {\cal O}_{6} \right|_{VS}
= - \frac{32\pi}{9} \alpha_s \langle \bar{q} q\rangle^2 
\simeq -0.0013~{\rm GeV}^6 \, ,
\end{equation*}
where we used $\langle \bar{q}q \rangle = -0.014$ GeV$^3$.  On the other
hand, analyses based on finite energy sum rules
\cite{DGHS,BGP,IZ,CAD+KS,Golo} typically find results
\begin{equation*}
{\cal O}_6^{V-A} = (-0.004 \pm 0.001) ~ {\rm GeV}^6 \, ,
\end{equation*}
which are not inconsistent with our number. The fact that we find
agreement within errors is not trivial. Since our approach is based on
less assumptions we may conclude that the sum rule results with their
relatively small errors are indeed trustworthy.


\section*{Acknowledgment}

We are grateful to A.\ H\"ocker for providing us with the ALEPH data and
for helpful discussions.




\end{document}